**First Draft on the xInf Model for Universal Physical Computation and Reverse Engineering of Natural Intelligence**

Last update: 2017. 12. 27. 6:56:17 UTC+8

*Hongbo Jia[1,2,3,4,5]

1 Suzhou Institute of Biomedical Engineering and Technology, Chinese Academy of Sciences
No. 88 Keling Road, Suzhou New District, Suzhou 215163, China. Email: jiahb@sibet.ac.cn
2 Institute of Neuroscience, Technical University Munich
Biedersteinerstrasse 29, 80802 Munich, Germany. Email: hongbo.jia@lrz.tum.de
3 Brain Research Center, Third Military Medical University
No. 30 GaoTanYanZheng Street, Chongqing 400038, China
4 Beijing AboroTech Co., Ltd.
Building D-807G, No. 2 ShangDiXinXi Road, Beijing 100085, China.
5 Chongqing U-LaborTech Co., Ltd.
Building 5-18-2, No. 101 CuiBai Road, Chongqing 400084, China.

* to whom correspondence shall be addressed, to either of the 5 affiliations with email addresses and/or postal addresses.

***Turing Machines*** are universal computing machines in theory[1]. It has been a long debate whether Turing Machines can simulate the consciousness mind's behaviors in the materialistic universe. Three different hypotheses come out of such debate, in short: (A) Can; (B) Cannot; (C) "Super-Turing[2]" (= "Hypercomputation", regardless whether being realistic or not[3])" machines can. Because *Turing Machines* or other kinds of theoretical computing models are abstract objects while behaviors are real observables (physical), this debate involves at least three distinct fields of science and technology: physics, informatics (computer engineering), and experimental neuroscience. However, the languages used in these different fields are highly heterogeneous and not easily interpretable for each other, making it very difficult to reach partial agreements regarding this debate, Therefore, the main goal of this manuscript is to establish a proper language that can translate among those different fields. First, I propose a theoretical model for analyzing how theoretical computing machines would physically run in physical time. This model, termed as the ***xInf Model***, gets its name from "arbitrary topology (***x***)" and "infinity (***Inf***)". I shall then show that this model is at first place Turing-complete in theory, and depending on the properties of physical time, it can be either Turing-equivalent or Super-Turing in the physical universe. The *xInf Model* is demonstrated to be a suitable universal language to translate among physics, computer engineering, and neuroscience. Finally, I propose a conjecture that there exists a *Minimal Complete Set* of rules in the *xInf Model* that enables the construction of a physical machine using inorganic materials that can pass the *Turing Test* in physical time. I cannot demonstrate whether such a conjecture to be testified or falsified on paper using finite-order logic, my only solution is physical time itself, i.e. an evolutionary competition will eventually tell the conclusion.

1. **Relativistic spacetime and causal order relation**

    Let us start with the most fundamental concepts of the physical computing machine, i.e. space and time – or in short, just one word "*spacetime*". However, we now deal with some different space: instead of the 3-dimensional observable physical space, we construct an abstract "*state space*" ***S*** of the computational states of the machine. Now let us try to include the *physical time **T*** into ***S***, and we encounter a problem: the physical computing machine must occupy certain volume in the 3-dimensional observable physical space, and thus it could be possible that the states in different parts of the machine do not have real causal order relation due to relativistic effect[4]. Ergo, a universal physical computing machine operating in physical time cannot be easily characterized by a simple topological *state space*. My solution to this problem is to split (reconstruct) the machine into countable (that can be either finite or infinite) number of components, each of which is sufficiently small and/or so wisely designed such that there is no relativistic effect to its physical computational states therein (Fig. 1a) and thus having 1-dimensional local time line. Each of such component is called a **M-Cell** (short for "Machine Cell").

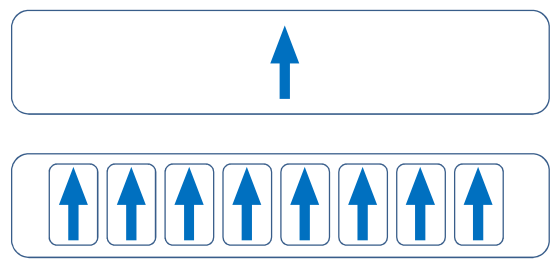

**Fig. 1a.** Construction of universal computing machine by using M-Cells. Upper graph: a computing machine; lower graph: a computing machine consists of multiple M-Cells. Each arrow illustrates a *time line*.

Due to relativistic effect, there is no guarantee that global causal order relations between states of different M-Cells can always be established[5]. Therefore, the entire machine may run without order, i.e. in chaos (= despite that each individual component has strict rules and orders, the behavior of the system cannot be described by orders). However, since that the computing machine can be controlled (programmed), depending on what specific *Rules* being applied onto it, partial order among M-Cells as well as complete order of the system can be established, such rules for distributed systems was pioneered by Leslie Lamport[5] and many people have contributed later on, e.g., Matherat and Jaekel[6]. Let us not address the issue of chaos and order here, first we shall have a mathematical notation system to establish the basis of discussion.

2. The xInf Model in math: relation to Turing Machines

In this chapter I shall give some mathematical descriptions about a physical computing machine as well as to define basic operating rules of those machines. The rules are arbitrarily defined without any knowledge *a priori*, i.e. these are axioms which one must accept for the moment and can be only testified or falsified by experiments.

Consider a given M-Cell named "*J*" within a physical computing machine (denote by "*Machine*" herein after). Since any M-Cell must operate by observables that can be measured at finite precision, e.g., electrical voltage of transistors or polarizations of photons, its computational states can be characterized by a *finite vector space* $V_j$ over a *finite field* $\mathbb{F}$. Let $T_j$ be the *time line*, the 1-dimensional space of physical time for *J*, defined over $\mathbb{R}$. The direct sum of $V_j$ and $T_j$ makes a new vector space, i.e. the *spacetime* $X_j$ of *J*

$$X_j = V_j \oplus T_j \quad \text{(Eq. 1a)}$$

To characterize the interaction between M-Cells as well as between a M-Cell and the Observer, let us decompose $V_j$ into three *orthogonal subspaces*, the *input space* $I_j$, the *output space* $O_j$, and the M-Cell's internal *state space* $C_j$, such as $V_j = I_j \oplus C_j \oplus O_j$, and rewrite the formula above as

$$X_j = I_j \oplus C_j \oplus O_j \oplus T_j \quad \text{(Eq. 1b)}$$

In practice one can consider the output space and the internal state space as one, denoted by $P_j = C_j \oplus O_j$ (*P* for "Program"), and we have

$$X_j = I_j \oplus P_j \oplus T_j \quad \text{(Eq. 1c)}$$

Computation is transformation of states in *spacetime*. A *Transformation Function* of a M-Cell, is a function $I_j \oplus P_j \oplus T_j \xrightarrow{TF} P_j \oplus T_j$ that satisfies absolute time order as:

**Rule #1 (Rule of Absolute Causal Order), state transformation within a M-Cell can and <u>must</u> occur with a non-zero, positive and finite time lapse in its *time line*.**

$$\forall t_1 \in T_j,\ I_j(t_1) \in I_j\ \&\ P_j(t_1) \in P_j;\ \exists! t_2 \in T_j\ \&\ P_j(t_2) \in P_j:\ TF\big(I_j(t_1), P_j(t_1)\big) = P_j(t_2)\ \&\ t_1 < t_2$$

Alternatively, one can also combine the input space and the output space as one or combine the input space and the internal state space as one, depending on how one would like to visually illustrate the M-Cell on paper by 3-D graph, i.e. let $E_j = I_j \oplus O_j$, $R_j = I_j \oplus C_j$, then we have

$$X_j = E_j \oplus C_j \oplus T_j \quad \text{(Eq. 1d)}$$
$$X_j = R_j \oplus O_j \oplus T_j \quad \text{(Eq. 1e)}$$

The entire *Machine*'s *spacetime* $X$ can be then expressed by the direct sum of all its M-Cells' *spacetime*

$$X = \sum_j X_j \quad \text{(Eq. 2)}$$

Now we come to the notion of connectivity. Consider a *Machine* with two M-Cells *A* and *B*, a *Connection Function from A to B*, is a function $O_a \oplus T_a \xrightarrow{CF} I_b \oplus T_b$ (Fig. 1b) that satisfies relative time order as:

**Rule #2 (Rule of Relative Causal Order), events of state information transmission from one M-Cell to another M-Cell must obey relative order, i.e., an event cannot arrive earlier than its previously transmitted event.**

$$\forall t_{a1}, t_{a2} \in T_a,\ O_a(t_{a1}), O_a(t_{a2}) \in O_a;\ \exists! t_{b1}, t_{b2} \in T_b\ \&\ I_b(t_{b1}), I_b(t_{b2}) \in I_b:\ CF(O_a(t_{a1})) = I_b(t_{b1})\ \&\ CF(O_a(t_{a2})) = I_b(t_{b2});\quad t_{a1} < t_{a2} \Rightarrow t_{b1} < t_{b2}$$

**Note: A M-Cell can also make a *Connection Function* to itself.** (Fig. 1c)

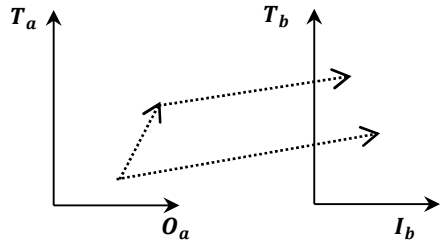

**Fig. 1b.** Connection function between M-Cells

In the case of multiple connections, i.e. multiple M-Cells connect to one M-Cell or one M-Cell connects to multiple M-Cells (Fig. 1c), one must further divide the relevant *input space* and *output space* into orthogonal *subspaces* and then establish the relevant connection functions, such process is called *Compartmentation* and *Connectivity*.

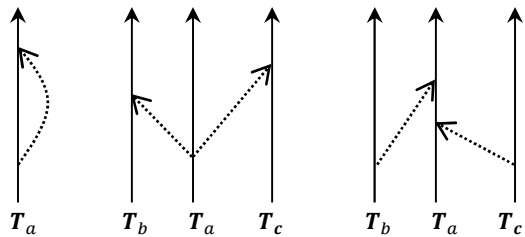

**Fig. 1c.** Possible configurations of inter M-Cell connectivity. Left: one M-Cell connects to itself; middle: one M-Cell connects to two others; right: two M-Cells connect to one.

State transformation and transmission events can be illustrated in a 3-D graph (Fig.1d) whereas operations of the *Machine* can be characterized by *Traces* in *spacetime* like the *world line* as in the *Minkowski Diagram*. A *Trace* is a strict *ordered set* of points in the *spacetime* $X_j$ of a M-Cell (a subspace of the *Machine*'s *spacetime* $X$) that satisfies strict order in $T_j$.

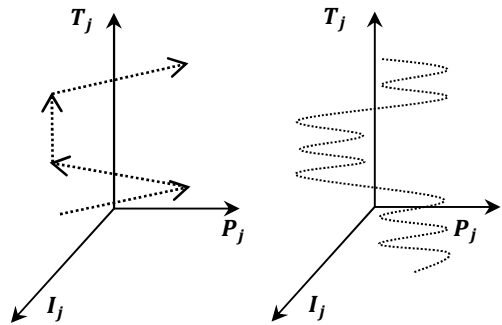

**Fig. 1d.** Examples of M-Cell's *Traces* in *spacetime*. Note that the right panel involves smooth curves but in principle it must be discrete like shown in the left panel.

There are a few differences though. First, the multi-dimensional computational state space replaces the physical space; second, the speed of light (information transmitting velocity) is arbitrary; third, state transformations occur as discrete events with respect to time. For any M-Cell, state transformation cannot happen if there is already a transformation underway, as in the following rule:

***Rule #3 (Rule of Trace Unity)***, **each M-Cell has and <u>just has</u> one *Trace* of state transformation in its *spacetime*.**

Note: this rule is against the fundamental concept of *Quantum Mechanics*; but we shall handle this issue in chapter 3.

A *Machine* must have at least one M-Cell to receive physical inputs from the *Observer* and at least one

M-Cell to transmit computational results to the *Observer*, denoted by two sets, the *Input Cell Set I*, and the *Output Cell Set O*, respectively. Let $i \in Z_i$ and $k \in Z_k$ be the indices of those M-Cells belonging to *I* and *O* respectively. Let $I_M = \sum_{i \in Z_i} I_j \oplus \sum_{i \in Z_i} T_j$ and $O_M = \sum_{k \in Z_k} O_k \oplus \sum_{k \in Z_k} T_k$ be the *input spacetime* and the *output spacetime* of the *Machine*, respectively.

Note that the *Observer* also has its own local *time line*, which is independent from those of the M-Cells. Furthermore, the *Observer* may also be a *Machine* that has its own M-Cells. However, we can always define a time dimension $T_g$ as the *global time line* for information exchange between the *Observer* and the *Machine*. Input events and output events occurring along this *time line* must have certain order along $T_g$, therefore, we can define a *global input space* $I_g$ and a *global output space* $O_g$ as the vector space of all possible input states and the vector space of all possible output states to direct sum of the *global time line*, respectively.

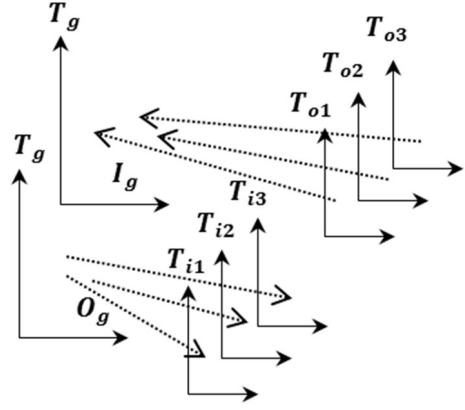

**Fig. 1e.** Illustration of the interface between the *Observer* and the *Machine*.

Note that both $O_g$ and $I_g$ belong to the *Observer* and must be *isomorphic* to $I_M$ and $O_M$ of the *Machine*, respectively. Whereas the functions relating them are called the *input interface functions* and *output interface functions*, respectively: $I_M \stackrel{IIF}{\leftrightarrow} O_g \oplus T_g$, $O_M \stackrel{OIF}{\leftrightarrow} I_g \oplus T_g$, as illustrated in Fig. 1e.

It is possible that there are multiple *Observers* that interface with one *Machine* and each of them has its own *time line* and *interface functions*. For example, two humans can talk to each other and being *Observers* for one computer, likewise, two computers on internet can communicate with each other while interfacing with one human (being *Observers* for the human). In this view, *Observers* are equivalent to each other, can interface independently with the *Machine* by their own *global time line*, *input spacetime* and *output spacetime*. However, *Observers* of the same *Machine* are not allowed to exchange with each other regarding any information not belonging to the specific input and output spaces of each interface.

People may find out that I introduced a *duality* of time: the time that makes the *Machine* working (so to simulate natural objects) and the time of the *Observer* that makes it perceive what does the *Machine* work out. Indeed, this duality is in line with the concepts of Sir Isaac Newton back to the 17[th] century[7]. However, an *Observer* can also be a *Machine*, regardless being biological or non-biological, we shall discuss about this issue later.

Definition: A ***xInf Machine*** is an object in the physical universe that satisfies the following requirements:
(i) Has at least one M-Cell and *spacetime* $X$ defined as Eqs. 1a-1e & 2 that obey **Rule #1, #2 and #3**.
(ii) Is interfaced by at least one *Observer* that has its own independent *time line*, *input spacetime* and *output spacetime*.
(iii) Obeys a set of *Rules*: regarding the following five categories as a **Minimal Set**:
(iii-1) *Birth* and *Death* of M-Cells (adding/removing an independent *time line*)

(iii-2) *Compartmentation* and *Connectivity* of M-Cells (adding/removing and partitioning of state space dimensions with respect to each *time line*)

(iii-3) *Transformation Functions* of individual M-Cells

(iii-4) *Transmission Functions* between pairs of M-Cells

(iii-5) *Interface Functions* between *Observers* and the *xInf Machine*

A *Minimal Set* guarantees the operation of a *xInf Machine* in the physical universe with physical *Observers*, however, it does not guarantee its computability i.e. what it can compute or simulate. For any physical or theoretical objects e.g. an apple, a star, a magnet, an electron, a Turing Machine, a qubit, a superstring, or a spin foam, etc., if there exists a set of rules for a *xInf Machine* to completely simulate it, such a set is called a **Complete Set** for the simulation of this object. Note that it is possible that for the same object there are multiple different *Complete Sets* to simulate it by *xInf Machine*.

Out of all those objects, I shall give one example: the *Turing Machine*. There are two (and more) ways to demonstrate.

(A) "Direct, physical simulation". Let a series of M-Cells $X_j$ play the role of each cell of the tape of the *Turing Machine*, one M-Cell $X_{head}$ be the head including its state register, and one set of M-Cells $X_i$ operating as the instruction table, with connectivity diagram as shown in Fig. 1f. The operating of the *xInf Machine* can exactly follows how the target *Turing Machine* operates in real-time.

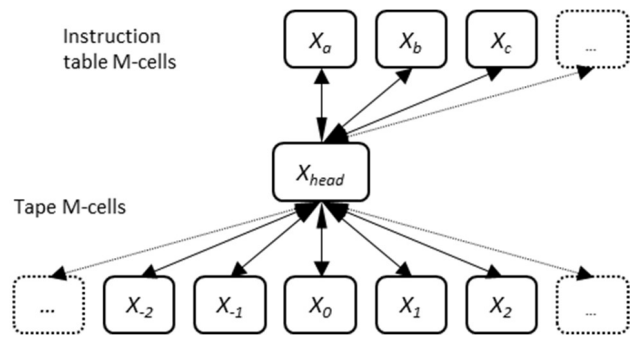

**Fig. 1f.** Simulation of *Turing Machine* by using *xInf Machine*.

(B) "Indirect, theoretical simulation". Construct a *xInf machine* whose structure and operations are identical to Stephen Wolfram's "Rule 110 Cellular Automaton"[8] as shown in Fig. 1g. Since this automaton has already been proven Turing-complete[8,9] we easily prove our *xInf Machine* to be Turing-complete.

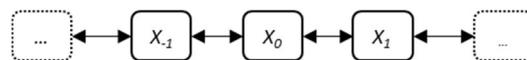

**Fig. 1g.** Simulation of Wolfram's Cellular Automaton by using *xInf Machine*

The major difference between *xInf Machine* and *Turing Machine* is just about one thing: time. As defined, the *xInf Machine* never halts (unless it runs out of physical energy, but this issue shall not interfere with computability), whereas "halt" is a state or part of a state that lasts for certain duration of time in the *output spacetime* of it. The only problem is that for infinite number of M-Cells the physical space required also becomes infinite and thus the transmission time for the far-away M-Cells to become infinitely long. However, since we are considering a theoretical *Turing Machine* that has infinite space, we are also allowed to consider something else in theory, consisting of finite space boundary and infinite number of *time lines*. Note that the notion of infinity is about the number of dimensions, not about the metric (distance).

The second way of proving involves *Cellular Automata*, in fact the *xInf Machine* can be regarded as more generic version of the original *Automata* of John von Neumann[10]. Two major differences shall be mentioned between the *xInf Machine* and the *Cellular Automata per se*. First, M-Cells in the *xInf Machine* update their states along their own *time line* independently from their adjacent M-Cells; second, M-Cells connect not just to their neighbors but also to any other M-Cell in the same machine whenever needed, the connectivity topology of a *xInf Machine* could be the same as a *Cellular Automata* but could also be fundamentally different.

In the physical universe, both *Turing Machine* and *xInf Machine* must operate in finite space and finite time. I have shown that *xInf Machine* can simulate *Turing Machine* in *real-time*. What about the other way, can *Turing Machine* simulate *xInf Machine* in *real-time*? Consider a *Turing Machine* to be a special *Observer* of the xInf Machine, that it can register all possible state transformations of all M-Cells as well as all possible inter-M-Cell transmissions. This *Turing Machine* is certainly capable of simulating that *xInf Machine*.

However, since *Turing Machine* has only one *time line*, i.e. the *Observer*'s *time line* $T_g$, to register all those above-mentioned events there must be projection functions from the *time line* of each M-Cell $T_j$ to the *global time line* $T_g$. Now we come to a very hard question: is the physical time discrete or continuous? (A) Continuous; (B) Discrete with a minimum quantum; (C) Discrete with arbitrary quanta. I do not expect any definite answer to this question from anyone or myself, however, let us see what will option (B) bring us. Let $\epsilon t$ be the minimum quantum of physical time, for example, the Planck time 5.39121 x 10$^{-44}$ s or whatever non-zero but constant value, meaning that any state transition in a physical *Turing Machine* cannot happen in time intervals shorter than it; for each M-Cell, there exists similarly a minimal interval time denoted by "M-Cellcycle" which is the minimally required time step $t_j$ as a minimum cycle time for any computational state transformation to happen in M-Cell *J*, which by definition must be greater than $\epsilon t$ : $\forall j, t_j > \epsilon t$; now that each M-Cell has its independent *time line*, the minimal time interval $\Delta t$ that any computational state transformation may occur in the *xInf Machine* (ensemble of the M-Cells) is the greatest common divisor (gcd) of all $t_j$ :

$$\Delta t \leq \gcd(\{t_j\})$$

even for realistic values that are far larger than the Planck time such as a few milliseconds of the refractory time for a neuron to pause between two consecutive action potentials firing, it is not difficult to conclude that one can always construct a *xInf Machine* with combinations of $t_j$ that will result in $\Delta t$ to be smaller than any defined $\epsilon t$; ergo, no *Turing Machine* can simulate this *xInf Machine* in *real-time*.

As shown in the demonstration above, the only chance for *Turing Machines* to be able to simulate *xInf Machines* in *real-time* is that physical time is either continuous or discrete with arbitrary quanta such that the *Turing Machine*'s operating step time can be arbitrarily smaller than the greatest common divisor of all M-Cellcycle $t_j$ of the *xInf Machine*. Without prejudice, I personally vote for the option that physical time is discrete with arbitrary quanta. In this context, *xInf Machine* is Turing-equivalent to *Turing Machine*.

**3. The xInf Model in Physics: relation to quantum mechanics and spacetime**

First, I would like to mention one famous figure – Sir Roger Penrose. He may not be the first one who tried to unify the two core theories in physics - *Quantum Mechanics* and *General Relativity* but he is most likely the first human who openly made an effort to find a trinity, i.e. understanding *Quantum Mechanics, General Relativity* and *Consciousness Mind* together by one unified theory[11]. Unfortunately, the model proposed by him together with Stuart Hameroff named "Orch-OR" (Orchestrated Objective Reduction[12]) have not managed to convince the majority of audience. Over a long course of debate[13], counterforces initiated by the calculations of Max Tegmark[14] have faithfully demonstrated that the hypothesis regarding microtubules' quantum entanglement states in neurons is falsified[15] and quantum computations in brain are very unlikely to happen[16], while Penrose and Hameroff did their best to defend their postulates[17]. In addition to those direct criticisms and arguments I would like to mention the fact that there are quite a lot of other kinds of molecules attached to and move over the microtubules[18]. The roles of such molecules are primarily for transporting other molecules that are required to maintain basic life and signaling functions at each part of the cell. Therefore, I would like to argue that, with whatever model to calculate the states of the microtubules there is no way to neglect the existence and impact of those motor molecules driving along them. The same ideology can be found throughout this manuscript, that **a** **good theory cannot rely on just a few arbitrarily selected elements while neglecting others that have strong interactions at the same spatial and temporal scale**.

Nevertheless, the courage and desire of finding such a *Trinity Theory* still stands for many physicists. With this desire let me now address the potential concern of physicists: suppose that there is a "*Quantum Computer*"[19], that can transform all states simultaneously at the same time, all of my previous postulates would become void. I cannot address quantum computation here because I have done nothing and know almost nothing in this field. However, my wife, my PhD supervisors and I together do have seen something that behaves like a multi-bit quantum computer by our own eyes via the measurement machine that we built. It was a cortical neuron in an intact living mouse brain with 50 (and more) synapses that we could observe by using a special microscope called the two-photon microscope[20] that was fast and sensitive enough[21] to capture the transient rising of fluorescence signals induced by calcium ion influx during synaptic activation[21-23]. Those independently and stochastically activating synapses behave very much like (entangled) qubits, whereas upon each time of sensory stimulation event the activation pattern comes out in highly variable yet distinguishable modes[24]. Let me rehearse that paper by re-arranging the original figures and texts plus some interpretations for those who are not familiar with neuroscience, with permissions (http://www.pnas.org/site/aboutpnas/rightperm.xhtml).

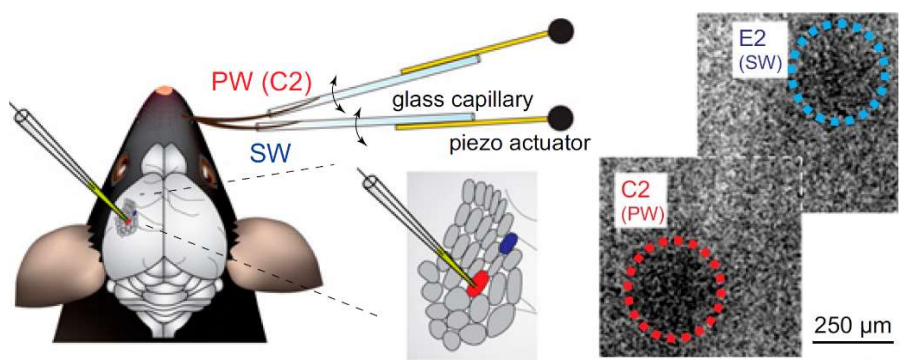

**Fig. 2a.** The two-whisker experiment design. Note to physicists: we are not dealing with the up-down spins of positron-electron pairs or the polarization directions of photons here, we are dealing with two whiskers (out of many) of a mouse, the mouse brain is capable of sensing which whisker is wiggled so these whiskers are physical *observables* for the mouse brain, readers who do not have experience with whiskers please wiggle your longest moustache hair and feel it (although the mouse whiskers are very different from human hairs, for a mouse they are more like fingers for a human, according to Bert Sakmann the 1991 Nobel Prize Laureate). Left cartoon picture shows a mouse head with two whiskers spared and attached to piezo actuators (other whiskers acutely removed prior to experiment begin) and brain exposed, in practice, only a small craniotomy of about 1.5 mm x 2 mm was opened on the skull. The image on the right side was taken before the craniotomy was performed where the micropipette was inserted into the brain. This montaged image was taken by a camera and using illumination of a 625-nm red LED, this procedure was called intrinsic optical imaging, whereas the darker regions on the image indicate more light absorption by deoxygenated hemoglobin in microvasculature and thus indicate more cellular metabolic consumption induced by the relevant sensory stimulation (in this case, to wiggle the whiskers by driving the piezo actuators back-and-forth at 10 Hz for 1 second). Such a method has been since long pioneered by Amiram Grinvald[25] to create functional maps of the brain cortex for experimental animals (as shown in the middle-lower cartoon), in an analogy to the well-known fMRI mapping for human brains. Whiskers are named by their row and column numbers as those two kept in this experiment are C2 and E2, respectively. The measurement of a single neuron later (where a micropipette was inserted) happened in the cortical region mapped to the C2 whisker, therefore we renamed the C2 whisker as "primary whisker" (PW) and the E2 whisker as "secondary whisker" (SW) and marked them in red and blue, respectively. In the mouse brain there is a region

where those functional regions mapping each whisker together form a barrel-like structure, thus this cortical region is called the "barrel cortex". Note that the sensorial cortex has a certain thickness and consists 6 anatomically discriminable layers named as Layer 1 to Layer 6. The 6-layered structure of cortex is largely universal for all mammalian brains, rodents and primates alike, despite that in some occasions some layers are taken together or further divided. It is a bit of good fortune that the presumably most "intelligent" part of brain is near the outer surface of it, in the sense that primates have the thickest (relatively) Layer 2/3 among all animals[26]. In this experiment we were looking at neurons in the Layer 2 of barrel cortex column C2.

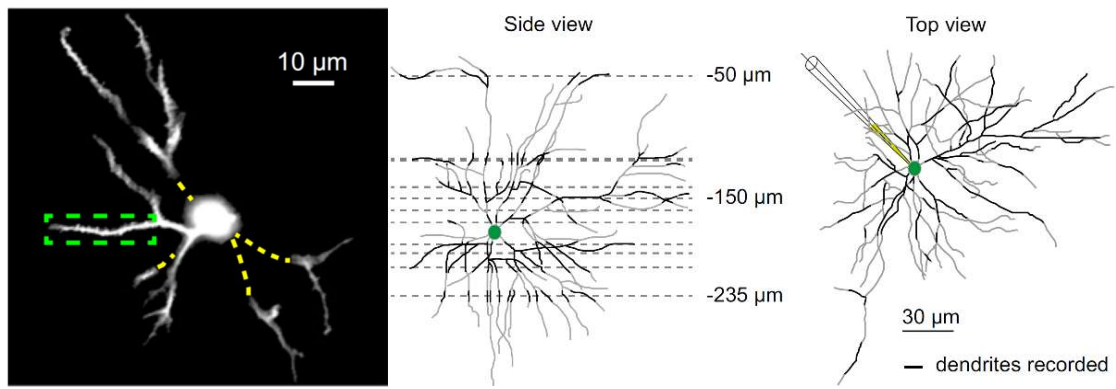

**Fig. 2b.** Two-photon imaging of a pyramidal neuron in Layer 2 of the barrel cortex column C2 in-vivo. Two-photon laser-scanning microscopy (TPLSM) is one special kind of optical microscopy technology developed by physicists[20] but it found its largest popularity among neuroscientists to be used for looking at cells in animal brains (in some occasions also in cut-out human brain tissues after surgery or postmortem donation), whereas four people among the inventors and users of this technology shared the 1-million-euro "Brain Prize" in 2015 (http://www.thebrainprize.org/flx/prize_winners/previous_prize_winners/prize_winners_2015/). But these experiments are not that easy to perform: it takes half a year or even longer for a well-educated people to be trained to practice the surgical and experimental procedures to achieve recording conditions and quality that are acceptable and repeatable[21]. One of the most difficult experiments is this one shown here in steps: (1) To use the microscope to navigate a micropipette with a tip of only a few micrometers through that small hole on the skull into a living mouse brain, while damaging as little objects as possible on the way, to find a neuron in the cortex underneath the dura matter. (2) To attach the neuron and form a tight electrochemical link to the neuron by means of having electrical resistance less than 50 MΩ to the interior medium of the neuron but more than 500 MΩ to the outer liquid environment. There is enormous amount of skills and experiences involved here, this technology named "patch-clamp" by itself[27,28] earned a Nobel-Prize for Erwin Neher and Bert Sakmann in 1991. (3) The most difficult part – to maintain this configuration for long enough (> 1 hour) so to have enough time to do these many things or let things happen: let the $Ca^{2+}$-sensitive fluorescence dye solution in the micropipette diffuse into the neuron and fill the fine stretches where the neuron receives synaptic inputs from many other neurons (called dendrites) so to be able to see the neuron's activity by means of calcium ion influx that is sensed by the $Ca^{2+}$-sensitive fluorescence dye. To wiggle the whisker of mouse and in the meanwhile record video-rate live image sequences of the neuron, this action needs to repeat many iterations because in each recording session one can only record from one horizontal optical focal plane (note: recent advances in imaging technology now allow simultaneous imaging of multiple planes or a certain volume) like shown in the left image, the dendrites appear as disconnected segments, and in the total duration of experiment only a fraction of all the dendritic tree can be recorded, like shown on the middle and right sketch drawing. In each focal plane, the stimulation of each whisker needs to repeat 10 trials (and more). In this configuration, we "hacked into" the "matrix" of the mouse, interfaced with one neuron while monitoring its connections to the "matrix" at the same

time when we do something on the outer materialistic world of the "matrix". Note: up to today there are some alternative and easier approaches for doing this kind of experiments as well. For example, one common approach is to use genetically-encoded calcium indicators[29] instead of the synthetic small-molecule dyes so to save the effort of micropipette manipulations.

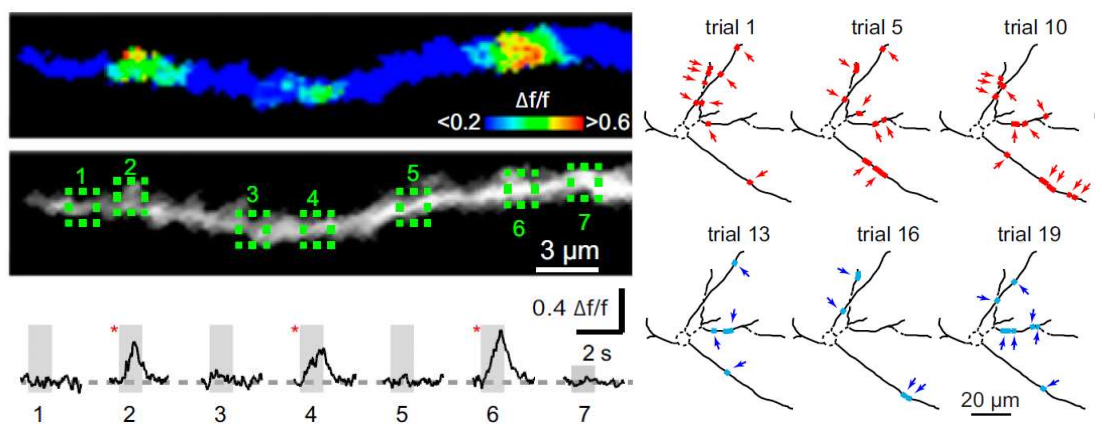

**Fig. 2c.** Preliminary analysis of dendritic hotspots. As one can see from the previous figure that there are multiple segments of neuronal dendrites and one of such segment is digitally magnified as in the image on the left side. Upper-left is a pseudo-color image that represents the relative fluorescence signal increment within 1-second integration time after one whisker stimulation event. This pseudo-color image shows a few "hotspots" in space, if we look in the time dimension as the lower-left traces, those 7 traces indicate the fluorescence change over time course for each of the 7 marked positions (greed dashed box on the middle-left image). These hotspots were known to represent incoming synaptic transmissions from other activated neurons to this neuron at these places along its dendrites[21,23]. The interesting point is that over repeated trials of whisker stimulation, the spatial pattern of activated hotspots vary from trial to trial, as shown on the right sketch images. However, readers can easily tell by eye that the red-marked hotspots (C2-whisker evoked) and the blue-marked hotspots (E2-whisker evoked) seems to be distinguishable to some extent, that we shall see in the next figures. There is a little trick involved in this measurement and analysis to acquire this kind of data. First, let me elaborate that the molecular signaling in a real neuron is by far more complicated than an artificial neuron. In an artificial neuron, an input is an input, an output is an output, they never physically or digitally interfere each other. But in a biological neuron (for most kinds of those in mammalian brains) there is a fundamental feature that the output spike event is transmitted back along dendrites to most sites of synaptic inputs and interact with them, such a phenomenon is called "back-propagation"[30-33]. This is also a typical example of conceptual translation between neuroscience and computer engineering, whereas in artificial neural network theories the same word "back-propagation"[34,35] has a different mathematical definition and underlying mechanisms. Without getting much into a new debate, I just mention that in neuroscience experiments it is not easy to tell apart whether the observed signal represents purely the synaptic inputs or is a mish-mash of both the synaptic input and the action potential output because the typical measurement precision does not allow reliable temporal segregation of the two (within few milliseconds' interval). Therefore, the micropipette (which was the most difficult part of the experiment) played a critical role here, by clamping the electrical current flowing through it to and from the neuronal cell body, we were able to suppress the neuron's action potential firing. This method is called "hyperpolarization"[21] (physicists please do not confuse with the polarizations of photons), that some neuroscientists have been skeptical of such trick since they believe that synaptic inputs over the dendrites are also affected. However, the current-clamp to the cell body does not affect much synaptically induced $Ca^{2+}$ influx at dendrites because the majority of $Ca^{2+}$ influx occurs in individual dendritic spines that are electrochemically independent compartments from the cell

body[36].

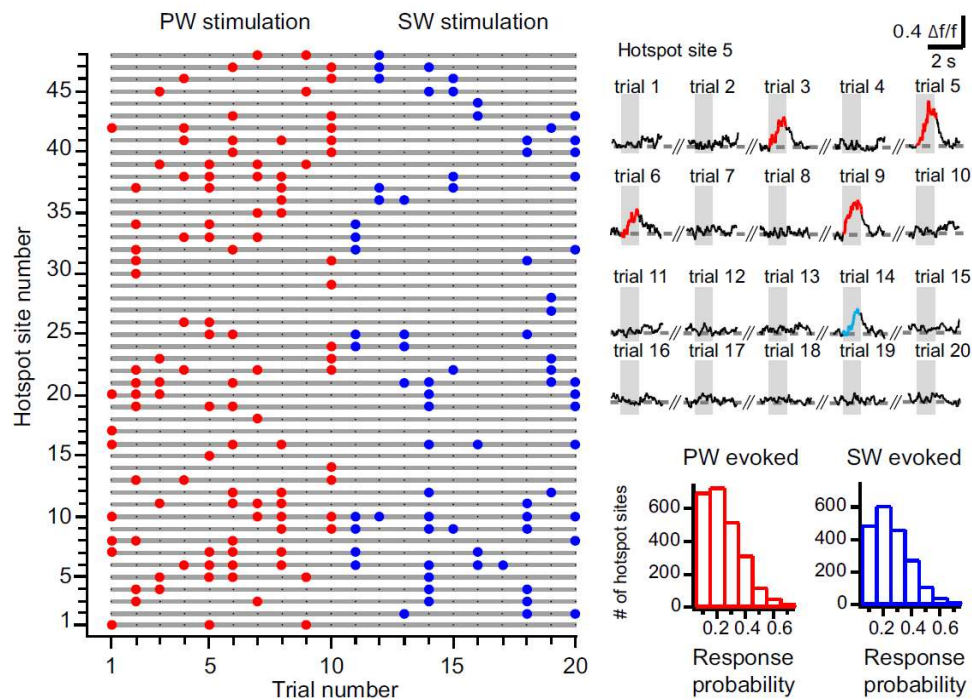

**Fig. 2d.** The dendritic input codes. The abacus chart on the left shows one full session of experiment (different from in the above figure). Please read this abacus graph as following: horizontal axis: time (discrete, each point means a random time interval after the previous one, of a few seconds or so), on each time point there is a whisker stimulation event. Horizontal axis: space, each row belongs to a spatial location along the dendrites. All red colored dots are induced by PW (C2-whiser) stimulation and all blue dots are by SW (E2 whisker), respectively. The same color representation scheme applies to the entire data analysis. Note that the repeated stimulation events were delivered at random time intervals, so that the traces for one single hotspot site are shown with the "//" label between, whereas the little gray column shade marks the stimulation event's time span (1 second). The lower-right bar graph shows the mean response probability for the ensemble of hotspot sites. The Gaussian shapes (a bit skewed) of probability distribution resembles the stochastic nature of all the sensory input hotspot sites. So how do you interpret this experiment? First try to use your eyes to perceive it, not a 2-D matrix in computer programs. The nerves called oculomotors driving your eyeball movements are essential part of your visual perception. You have two ways, either to sweep first horizontally and then vertically, or to sweep first vertically and then horizontally. Move on to the next paragraph and figure after you have swept several times and got some impression.

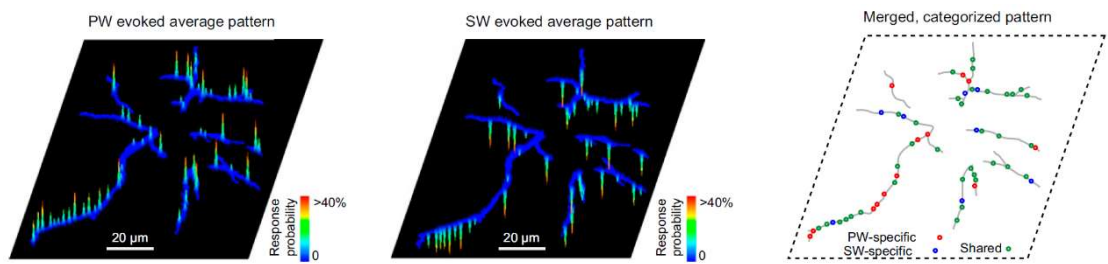

**Fig. 2e.** The spatial code. If you did the first-horizontal-then-vertical observation, i.e. to first integrate over time and then compare in space, you get the same kind of impression in this figure. The left two pseudo-color images show the average (=integrated, because the integration time is the same for every hotspot) response probability

at each spatial site of the dendrites for the two different whisker stimulations. The result can be further quantified into three categories of hotspots: PW-specific, SW-specific, and "Shared". This means, if you integrate long enough time, you can tell apart in each spatial location what kind of information in coming to the neuron.

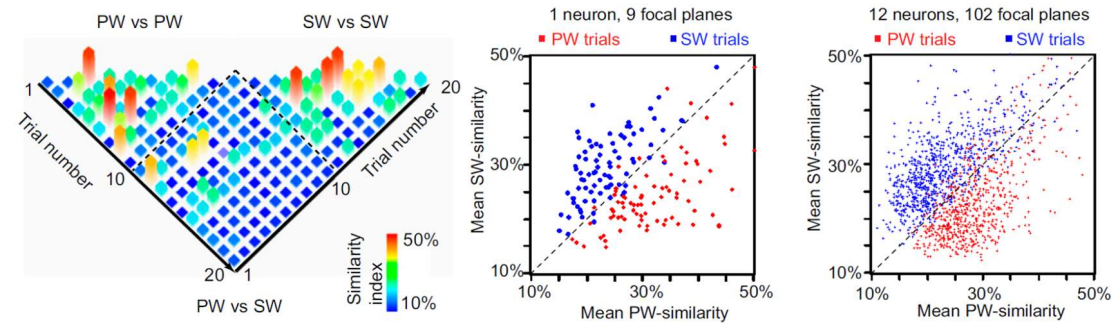

**Fig. 2f.** The temporal code. If you did the first-vertical-then-horizontal observation, i.e. to first make a pattern out of each stimulation event and then compare them using a simple similarity index, we get you this figure. The response patterns induced by the same whisker are more like each other than those induced by different whiskers. The left graph shows this similarity metric measurement in one recording focal plan in one neuron (10 trials of PW and 10 trials of SW stimulation), and the middle and right graph shows the accumulated data for all focal planes in all neurons. This means, despite that you are only looking at a small fraction of the entire dendritic tree of a neuron, if this part contains sufficient number of active hotspots, you can tell apart what time it was (what whisker being wiggled). Sounds like holography, right? But…

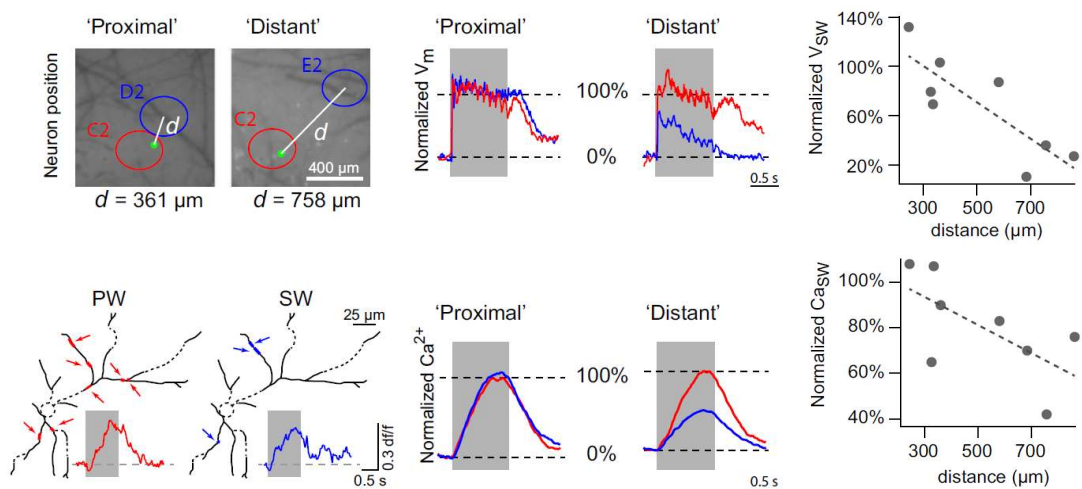

**Fig. 2g.** The integrated spatiotemporal code in the neuron. No matter how fancy pattern recognition algorithms you can perform with these data, they mean nothing to that neuron being poked. The neuron only has one output time line, i.e. at a place under the "butt" of its cell body there is a part of organelle called the "axon hillock", it is the place where the action potentials are initiated and propagated along the root-like structure called "axon" to deliver neurotransmitters via synapses to other neurons. And so is how the natural neural network working from a neuroscientist's point of view. The action potential firing by itself is a stochastic process too, involving complex electrochemical mechanisms, but it is well known that the firing probability is positively correlated with the integrated membrane potential level shown as "Vm" in the upper middle chart. The upper left image shows different experiments whereas the SW was assigned to other whiskers so to have difference distances of the SW-barrel column to the recorded neuron. For example, the center of the D2-barrel is much closer to the recorded neuron who sits in the C2-barrel, so this configuration is called "proximal" while the E2-barrel is further away so to be called "distant". This result agrees with other previous neuroscience findings

that the further away whisker (cortical barrel) induces less integrated response to the neuron. Very simple causal order, that's all. What a big chaos we had to make just to get this simple order!

Note that these segments of dendrites are just a small part of one neuron within a 1-million-neuron mouse brain, may I ask a simple and naïve question that what would quantum computer makers and quantum mind believers say about how brain works and whether their quantum computers will ever be able simulate the brain in this context?

Now let us come back to *Quantum Mechanics* itself. Since measurement itself is related to time, we can come up with a new *interpretation* of *Quantum Mechanics* called "*Real-time Interpretation*" with three basic concepts: no probabilistic wave functions, no instantaneity, no parallel universes. There is only one universe: the universe. Everything in the universe must be real, measurable by physical instruments (that are made by physical objects) in finite time intervals. So, how do we explain the *quantum uncertainty* experiments? Very simple and straightforward:

(i) Each object has its own *time line*(s), no matter being a brain, a neuron, an arm, a robotic arm, a block of iron, a star, an electron, a photon, a proton, a quark, etc.

(ii) The *Observer* has its own *time line* independent from those of the object. The observable (measurable) states are the output states of the object that can be transmitted to the *Observer*'s *time line*. Such transmissions for example can be: one single photon hitting a photodetector and inducing a transient electrical current; neutrinos from far-away galaxies hitting Cherenkov detectors, etc. Transmissions can also be: fluorescence photons from biological cells reaching microscope's image detecting system; photons emitted by display screens of measurement machines to the retina of eyes; acoustic waves from natural objects propagating to ears. Note that transmissions can be combined as well, just take the first two examples together and we get how the "Super-Kamiokande" project[37] basically works, where 11200 photomultiplier tubes (PMTs) have been used. This is also how our nervous systems can perceive and recognize much more that what our retina and cochlea can normally detect.

(iii) A measurement action forces the object to transmit its observable output state to the *Observer* as well as forces the object to receive an input state from the measurement device during the time course of measurement. Whatever results come to the *Observer* are determined by the output state of the object at the time of transmission, and whatever state transformations in the object are determined by the input from the measurement action and the object's own *spacetime* functions.

To make it clear, the above three rules of *Real-time Interpretation* does not interfere with existing established *Quantum Mechanics* theories, whereas all the calculations of probability distributions using whichever representation – Schrödinger, Heisenberg, Feynman, Penrose, etc. can still stand. The *Real-time interpretation* is just about the understanding of the measurement uncertainty and the quantum probability.

After all, I shall make a summary of the general principle: **by organizing the stochasticity and ordering in *time* together with rules of transformations and transmissions in *spacetime*, one can achieve desired stochasticity and ordering in *space* over a certain duration of *time*.** Time itself is the missing *hidden variable* of Einstein[38,39], although it is not really "hidden" – time is everywhere (and thus nowhere). The first (but incomplete) demonstration that hidden time as the hidden variable does not violate the *Bell's Inequalities*[39,40] was given by Pavel Kurakin[41]. Be advised, the word "time" has not been properly defined, neither in the history of physics nor in this manuscript. For the arguing issue of "***what is time?***" I have exchanged opinions with Marc-Thierry Jaekel[6,42] who is a professional physicist working at the École Normale Supérieure, Paris, and by general virtue of science I am allowed to cite his original text here:

> *…In your mail, you have acutely remarked that the existence of two different notions of time was pointed at*

*in the introduction of our work on concurrent computation with Philippe Matherat. This distinction has remained unnoticed for long but underlies the foundation of modern physics. Our comment in fact comes from carefully reading the 'Principa Mathematica' and acknowledging the consequences of Newton's work on the development of both relativity and quantum theory. It explains the deep antagonism between these two theories, and the origin of the main difficulty encountered by physicists to elaborate a consistent framework including both theories. This attempt has remained so far unsuccessful, and physics still follows a schizophrenic life: a mathematical, unobservable time, for describing the quantum evolution of atoms and particles, and a physical, observable time, for describing the spacetime positions of macroscopic systems. The use of two incompatible notions of time has been persisting since Newton's representation of the physical world. In our paper with Philippe, and in the following one on clockless circuits, we tried to show that the same duality (as you call it) affects computing devices, in the sense that concurrent systems rely on a notion of time which differs from that underlying ordinary computers, like Turing machines. The latter depend on the existence of a global time, determined by the clock, while the former can work without such a notion, but still depend on the existence of a time, which appears to be deeply rooted in physics, as it is necessary to ensure causality.*

*But I think that these two notions are both incorrect, or more precisely incomplete, in the sense that they are two approximate and simplified representations, preserving some properties at the expense of other ones, of the same concept. As you mention the particle-wave duality, let me just say that in this case the duality concept precisely showed to be a dead end, so that most physicists now agree that quanta are neither particles nor waves, but are an elementary concept.*

*The paper (by Kurakin) you put to my attention effectively seems to follow a similar line of thought. It focusses on a notion which is largely used in distributed computing systems, namely request-acknowledge links. These are necessary to ensure a global causality within an assembly of processors which are not linked to a common clock. This is a simple way to synchronize systematically without using a clock. In this way, one can imagine to immerse some clockless circuits into a global time and hence to emulate them on ordinary computers (Turing machines). But clearly (as can be seen in our papers), some clockless circuits use more complex ways to get synchronized. So that they use a reference to time (causality is preserved), but do not obviously share a global time. I think that the same critics apply to physical systems. Kurakin's example (the question of Bell's inequalities) besides the same defects, suffers from other ones. It focuses on a particular problem (related to EPR paradox), but in fact does not address it. EPR (Einstein, Podolsky, Rosen) raise the issue of the compatibility of quantum correlations with causality. Bell's inequalities assure that the correlations are genuinely quantum, and cannot be simulated by classical hidden variables. It is true that all theorems depend on assumptions, so that this does not prove that there cannot be a very subtle kind of hidden variables which turn out to satisfy these inequalities. But this does not mean either that such variables can be relevant for solving EPR paradox. Furthermore, Kurakin does not give any argument in favor of the existence of his 'hidden time' (the allusion to Feynman-Schwinger formalism is not conclusive). The problem with time is deep and general, and cannot be addressed on a single example.*

*As you know, I think that synchronisation must play a central role, and should be present in the definition of time…*

As an amateur physicist I do not have the "schizophrenic" feeling of the notion of time as professional physicists like Marc does. However, for many years I have been building and using measurement devices for many kinds of objects including natural neural systems (brains), with these devices my colleagues and I were able to observe how does a living animal (mouse) sense time[43]. If the time used to interpret quantum measurements is kind of "mathematical time" that is not perceivable, I shall mention that the time determining brain neuronal activity patterns is the perceivable physical time that by itself can be perceived by another objective entity (*Observer*). Taken together, I believe that **there is only one kind of time, i.e. the *physical time*, it makes objects** **containing information being transformed, transmitted, and observed. Physical time is distributed everywhere in space, and together makes non-separable *spacetime* unity. Whether the physical time (and physical *spacetime*) is arbitrarily partitionable or must consists of finite-sized quanta, is unknown, but it does affect the computability of *Turing Machines* in the physical universe**. The fact that any measurements of physical observables must occur in non-zero, positive and finite time lapse is in line with the notion of non-zero, positive and finite time lapse of state transformation in any physical measurement machine; moreover, any sensation on sensorial organs as well perception in the brain must also occur in non-zero, positive and finite time lapse. In this context, any of the "Gedankenexperiment (Thought Experiment)" if ever could happen, must also occur in non-zero, positive and finite physical time lapse, whereas those conceived instantaneous measurements can never happen in reality, let alone the brain perception of those Thought Experiments can never happen instantaneously.

There is indeed a phenomenal duality of physical time despite that the word "duality" itself has been discarded by the majority of physicists (based on Marc's summary, that I view as a measurement report of a population of physicists' brains). This duality of time, i.e. the time(s) of the object and the time(s) of the *Observer*, enables us as humans at space scale of a meter and temporal scale of some years to understand physics at all spatial scales within the same mathematical language framework (*Algebraic Geometry*), from elementary particles to microscopic organisms, and from daily living objects to astronomical objects. The only assumptions we need are the three basic rules about time given in chapter 2, i.e. the rules of absolute and relative causal order and the rule of trace unity.

The notion of different spatial scales brings forward an important issue of **spatiotemporal projection**. Such a technique was employed in the demonstration of using *Turing Machine* to simulate *xInf Machine*. The interface functions as mentioned to be rule set (v) are also projections functions if the number of involved interface M-Cells is more than 1. A projection function from multiple M-Cells to one reduces the number of dimensions in *time* to 1 but at the cost of increasing the number of dimensions in *state space* to the sum of all that of the projected M-Cells. Note that this projection can also work for infinite dimensions *state space*, i.e. the *Hilbert Space*. However, depending on the *spacetime* structure of the M-Cells there could also be degeneration of the dimensions of summed state spaces, i.e. using fewer *state space* dimensions to emulate the machine at certain precision (for the *Observer*). For example, to calculate the kinetic properties of a block of iron consisting of numerous (e.g. 1 Mole) Ferrum atoms one just need 6 spatial dimensions. A more complicated example is nanometer-sized materials that possess partially quantum properties and partially classical properties. This also explains why physical laws applying to different spatial scales have very different mathematical forms and in many occasions multiple forms for describing the same object, due to the differences in *spacetime* structures at those different scales such that the projection function can be ***approximated*** in multiple ways. Such projections from smaller to larger scales come out in four basic modes: (i) from order to order, classical physics works in this mode; (ii) from chaos to order, statistical physics as well as canonical quantum mechanics deals with such mode; (iii) from order to chaos, typically this is what the nominal *Chaos* theories are dealing with; (iv) from chaos to chaos, an example is what I will discuss in chapter 6, to emulate one chaotic system (brains) by another chaotic system (*xInf Machines*).

Furthermore, measurement of elementary particles must also involve spatiotemporal projection, as an example, photodetectors that are used to detect single photons must project the single photon's *spacetime* to many (thousands to millions) electrons' *spacetime* so to be able to drive electronic devices to record the detection event or to display the result by other many photons emitted from display screen so that human visual system can perceive and respond. Such a projection does not (and will never be able to) guarantee 100% reliability of reporting the single photon event (i.e., quantum efficiency) but does sometimes give readout even when there is no photon coming (i.e., dark counts), meaning both false positive and false negative events occurring at non-negligible rates. In fact, one of the most important things in practicing quantum communication is to develop single-photon detectors[44] that have high quantum efficiency and low dark count rates in combination with particular operating protocols of them.

Last but not least, there exists **certain *spacetime* organization that allows autonomous generation of partial order from stochasticity on a larger spatiotemporal scale at certain cost of physical energy dissipation** as well as forcing other objects to generate order from stochasticity or vice-versa. Some readers may already realize that an example of such organization is the **natural intelligence**.

Since long time ago there existed a shorter word to describe such phenomenon in many different research fields, i.e. "self-organization". Now we come from a different point of view, by establishing a new notion of *time* and categorizations of *spacetime* organization rules, we are now able to translate among those distinct research fields using the same mathematical language framework. But the fact is much more complicated: first, so far there are no other non-biological organisms beyond natural brains can self-organize to behave like what natural brains can do or just what *Turing Machines* can do. Second, even a biological brain is not guaranteed to be fully capable of conducting behaviors like what another brain or a *Turing Machine* can do. Something is missing here, so the word "synchronisation" in Marc-Thierry Jaekel's text above brings forward a hypothesis, that only in some particular modes with certain order relations within the brain or *Machine* can it sense physical time and thus to compute as a *Turing Machine* or perform logic of finite order. While he believes that it is neuroscientists' job to find out the rules that enables such modes, I think only the combination of both theoretical physicists and neuroscientists can achieve this goal. Ergo, a proper way of **translation** shall be established between the two distinct groups of people, which is one of the major points of this manuscript.

This is so far that I can do as an amateur physicist. Professional physicists and mathematicians among readers of this manuscript may soon find out that the *xInf Model* framework shown above can fit many concurrent theories in *Quantum Mechanics* and *General Relativity*, particularly for the *Loop Quantum Gravity* and those involving *spacetime* topology. Alternatively, they may also consider this model to be nothing but literal crap. Whatever decisions are nothing but state transformations in the *spacetime* of neurons in their brains after receiving the transmissions of these above texts and figures via light patterns and sequences to their eyes. But note my word that such decisions are changing over time, as the neuronal activity patterns are always going on, they never halt unless the cells are all dead. In contrary to the common sense that neurons do some kind of computations only when they are asked (stimulated) to do so, many labs have direct and abundant experimental observations that neurons undergo never-ending self-organized activity patterns called "spontaneous activity". Of particular importance, for dendritic spines of a single neuron in the mouse primary auditory cortex, it is the same set of spines being activated by sound stimulation as being activated during certain phases ("Up-states") of spontaneous activities[45]. Without going further into interpreting this and its related experiments, I just recall one famous quote from Einstein: *is the moon there when you are not looking?* With the same ideology of this manuscript I would answer "Yes, the moon is there, and not just being there, it is moving all the time when we are not looking at it, and we also know precisely at what probability it should be at this place and that time".

Therefore, I shall temporarily leave behind those works and arguments regarding existing objects and theories and now focus on the birth of objects that have not existed so far, i.e. the reverse engineering of natural intelligence. Let us move on to the next chapter of how to implement the *xInf Model* in digital electronic computing devices that are currently available in modern human industry.

**4. The xInf Model in Computer Engineering: Distributed Real-time Computation**

John von Neumann had established modern electronic computers' fundamental working model[46] named after him, i.e. the *von Neumann Architecture* (Fig. 3a) consisting of processor, input and output device, program-and-data (all-in-one) memory linked to a bundle of general connection lines called the *bus* (i.e. where data hop on and off). It is worthwhile to mention that the British *Colossus* machine was also kind of "*Stored-Program Computer*" that already exists and impressed him, according to recent history studies[47] involving declassified WWII documents. Nowadays the notion of stored-program computer refers to more kinds of computing machines, including the other most commonly seen "*Harvard Architecture*" machines (Fig. 3b) where the program memory and data memory are separated and handled differently. Modern computers have architectures that are more or less hybrid of both kinds, the only differences among those architectures are where and how exactly programs are stored, retrieved, and executed.

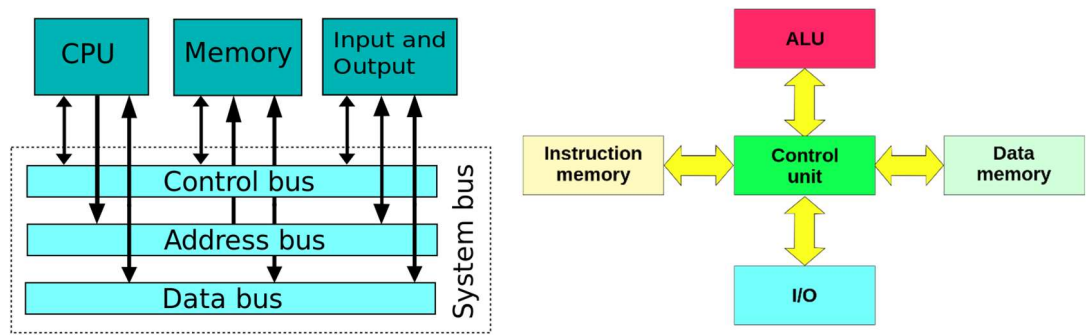

**Fig. 3a (left) & Fig. 3b (right)**. Illustrations of the Von Neumann Architecture (left) and the Harvard Architecture (right) of digital computing devices. It is also possible to think that the Harvard Architecture is not fundamentally different than the Von Neumann Architecture except that programs and data are handled differently. Images copied from Wikipedia under general licensing.

No matter how computers are constructed in history and nowadays, one thing is for almost 100% certain: that computers work very differently than natural brains. Von Neumann himself was the first to recognize and develop very profound thinking into this issue, illustrating in the book "*The Computer and the Brain*"[48] that was published after his brain stopped working. In recent years there has been rapidly increasing interest of many people to make computers work like brains, i.e. "brain-like computation". But what is the brain like? Does it look like a walnut? Or does it taste like Tofu?

In this chapter let us deal with computers and in the next chapter we shall deal with natural brains and how to make one behave like the other. On the contrary to stored-program computers, the *xInf Machine* does not have a stored program *per se*, i.e. no constant place to store any program. Program exists in every M-Cell but for the entire machine, there is no program because there is no global time order. Moreover, every function in every M-Cell must execute in at the correct time intervals within tolerable error range. Taken together, the precise engineering definition of a *xInf Machine* is "***Distributed Real-Time Computation System***". More importantly, this definition is also valid for natural brains. However, bearing the same literal definition does not necessarily mean the two objects are truly the same or similar, the only way to prove is by means of math and experiments.

I must emphasize once more the notion of ***Real-Time***: it does not mean how fast the computation could be done, it means that **every transformation function and every transmission function must apparently contain physical time and be executed in the specified time interval** (at the best possible precision), **running neither slower nor faster, neither earlier nor later, with neither interruption nor waiting for request and receiving** (if ever any time delay involves, that must be given with precise time values). Time is equivalent to data (space) in the *xInf Machine* in practical computation, errors in computing time inevitably lead to errors in computation results as seriously as for what errors in data would cause.

In fact, all stored-program computers are fundamentally "real-time" computers in the sense that every instruction is executed in defined number of clock cycles. However, it is the operating system which determines how "CPU time" is related to the physical time in the functions of M-Cell, because the instruction set is typically limited to just a few functions (albeit in theory *Turing-complete*). Thus, the compilation system and operating system determine in what order each instruction is executed so to implement complex functions in computation task, and the user (*Observer*) has to accept whatever time cost it takes to complete certain computation task. One can argue that the result of computation does not necessarily need to occur just at the exact time, either the *Machine* or the *Observer* can wait (for pre-defined time intervals) for the interaction of computational inputs and outputs. This argument is true for computation tasks such as playing *Chess*[49], *Go*[50] or *Poker*, pattern recognition of voices[51,52], faces[53] or medical pathological data[54] or prediction of stock markets, etc. However, it is not true for many other tasks such as playing the *StarCraft*[55], playing football game in a stadium[56], surgical operations[57] in soft tissues, or driving car[58] in normal city streets, etc. It is not difficult to imagine that in practical life, most tasks require non-pausing and intensive interactions that must happen at proper (but not necessarily pre-defined) time with proper (but not necessarily precise) result, otherwise critical failure or physical damage may occur, most of which are irreversible. Such tasks are typically considered as "natural intelligence" tasks which "artificial intelligence" (AI) is expected to be able to manage after certain time of development[59]. However, I must argue that the "AI" which will manage to perform such tasks would certainly not be that "AI" s as they are today, at least from hardware point of view.

Generally speaking, the implementation of M-Cell and *xInf Machine* turns out to be a composition of multiple "programmed processors" and "data memories" as shown in Fig. 3c. The only difference is to "solidify" every program onto processor in each M-Cell, and thus no operating system is involved. This would mean that **each "processor" in each M-Cell executes certain instructions in an endless looping manner at each own clock intervals** – until they break down or run out of energy. Such instruction sets fall into those five general subsets that I have listed in chapter 2 and are infinite in combinations, theoretically the *Machine* has infinite size of *instruction set* – that is where the name "*xInf Machine*" fits.

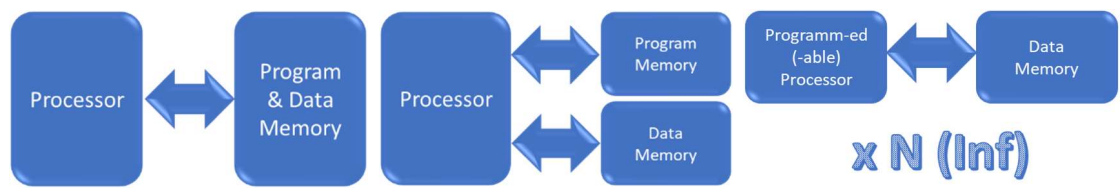

**Fig. 3c.** Comparison of basic computational architectures. Left: Von Neumann; Middle: Harvard; Right: xInf. The redrawn model graphs comparing to Fig. 3b shows that the Harvard Architecture and the Von Neumann Architecture may not be fundamentally different. However, the xInf Architecture is fundamentally different to the previous ones in the sense that there is no more "programs", they are moved onto (embedded into) the processors.

However, such an approach is quite difficult to be realized; in order to be able to implement millions and even more different kinds of programs (instructions) that can be executed in each own independent *time line*, millions of different kinds of processors (namely, application-specific integrated circuit, ASIC) have to be produced. Unfortunately, for modern semiconductor industry, despite that it is easy to produce millions and billions of pieces of any specific kind of ASIC, it requires significant effort and resources of humans to design so many different kinds of ASICs. Therefore, the solution is to change one word: from "programmed" to "programmable". Let the *Observer* be able to program the M-Cells and let some M-Cells be able to autonomously program the other M-Cells. The first half wish is readily realized by using Field-Programmable-Gate-Array (FPGA) as shown in Fig. 4a, while the second half wish needs some extra engineering effort.

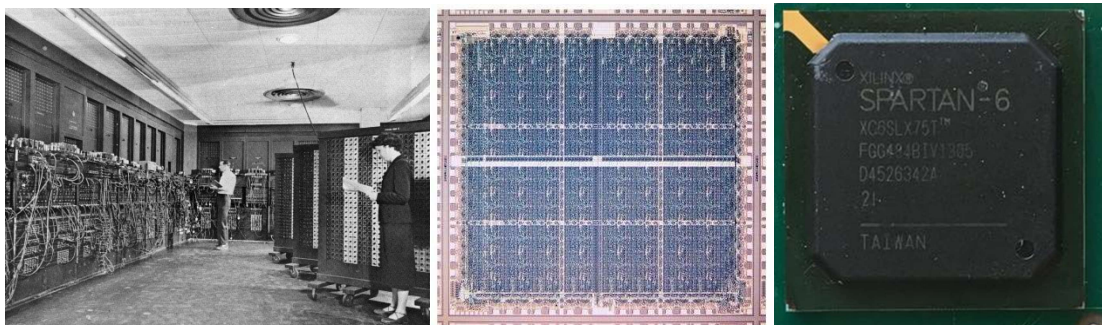

**Fig. 4a.** FPGA in history and modern times. Left: the first "Field" programmable gate array in history – Glen Beck (background) and Betty Snyder (foreground) programming the ENIAC in BRL building 328, photo and notes copied from Wikipedia. Note that the ENIAC and the EDVAC computers were fundamentally different, albeit both having been consigned by the United States Army and John von Neumann was involved in both. The EDVAC was a true stored-program computer but the ENIAC was effectively a FPGA except the "field" re-wiring was done by human hands instead of electronic signals. Middle: the first commercially viable FPGA chip in history, model XC2064 of Xilinx[TM]. Right: a low-cost FPGA chip that is used extensively for automatic instrument control, belonging to the SPARTAN-6 series of Xilinx[TM].

M-Cells shall be categorized into difference "cell types" such that the rules of how one M-Cell can change another M-Cell's program can be established (involving rules (iii), (iv) and (v) of the *Minimal Set*). Therefore, one would need to devise certain hierarchy similar to the biological cell genesis system, called the "M-CellGen" system. Features of this system shall include: (i) Unique and unclonable identity of each M-Cell hardware; (ii) Complete evolution history of each M-Cell's program can be kept and retrieved. (iii) Defined levels of programmable authority, that can only be changed by human programmers. Note that there is no necessity of having a global system of M-CellGen, so each individual person or organization can establish their own M-CellGen systems and rules. However, a global serial number mapping is still necessary when the M-Cells' hardware belonging to different people or organization are to be integrated together. This serial number system is more similar to conventional citizen ID numbering rather than to the internet IP address, because the devices to be numbered have freedom to form interconnected network with any arbitrary topology and do not need to subject to any regulation *a priori*.

One major problem has to be solved prior to directly connecting the M-Cells, i.e. as always – problem of the time: now that every M-Cell has its own digital clock(s), how to make sure that state information transmissions from one sender M-Cell can be 100% picked up by its receiver (and not like photodetectors which cannot have 100% detection efficiency)? The solution is provided by the pioneers of modern information and communication theory, Claude E. Shannon[60] and Harry T. Nyquist[61]: **Nyquist-Shannon Sampling Theorem**. Let us call the connection wiring and transmission protocols between two M-Cells as "M-Syn". As shown in Fig. 4b, each M-Syn must minimally contain two electrical signal lines, the first line for the sender clock and the second for the sender signal. Additional optional lines include energy transfer, signal ground, etc. The receiver M-Cell must receive the first two lines with a receiver clock at a frequency that is at least 2-fold higher than the sender clock frequency, i.e. oversampling. In this way, the signal processing component in the receiver M-Cell will be able to detect every tick of the sender clock and thus to acquire the incoming signal without information loss and transform them into a first-in-first-out (FIFO) to be used by the recipient M-Cell's state transformation program. Be advised, that multiple physical clocks being present in one M-Cell does not violate the definition of single *time line* per each M-Cell in theory, because these multiple clocks are synchronized with constant frequency and phase relations. In practical term, for example, a FPGA chip can use double-frequency of its onboard clock.

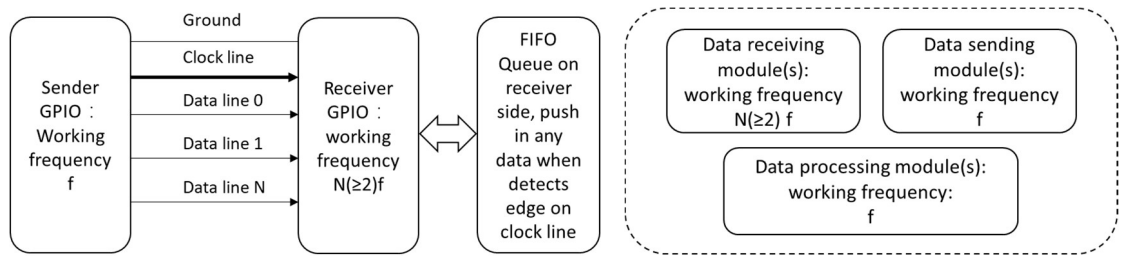

**Fig. 4b.** Illustration of the oversampling principle for data transmissions between M-Cells. Left: the oversampling protocol between two M-Cells. Right: relation of data receiving, processing, and sending on one M-Cell. Sending and processing data can use the same clock frequency while receiving can use higher clock frequency.

This general strategy of temporal oversampling is backed-up by experimental neuroscience findings: for a healthy neuronal synapse in healthy brain, the receptor molecules' refresh rates are faster than the release rates of neurotransmitter vesicles and the excess neurotransmitters are cleared out from the synapse by glial cells at faster rates than the average release rates, therefore the postsynaptic neuron can always pick up neurotransmission events from the presynaptic neuron. Note: the actual numbers of available receptors are varying over time so the response amplitude to each neurotransmission event is variable, but this is in the context of neuroplasticity that belongs to the category of transmission functions.

It is important to note that, the signal lines as mentioned above do not necessarily need to be digital lines. Analog signals of a limited bandwidth can also be transmitted via M-Syn. All together there are four basic subtypes of M-Syn, DD- (digital to digital), DA- (digital to analog), AD- (analog to digital) and AA- (analog to analog). In DA- and AD- subtype M-Syns the relevant digital-to-analog converter (DAC) analog-to-digital converter (ADC) must be present either in the M-Syn's signal processing component, respectively. So far, the only possible way for digital computing devices to interface with the rest of the physical universe is via analog signal converters. Analog signals do not necessarily require a clock, nevertheless, they could be considered as being accompanied by a digital clock running at the upper limit frequency of the analog signal bandwidth. For each M-Syn the receiver clock frequency $f_{receiver}$ and the sender clock frequency $f_{sender}$ must subject to the following oversampling rule: $f_{receiver} \geq 2 f_{sender}$

The practical engineering form of M-Cell (called "EM-Cell") may differ from the theoretical M-Cell in the sense that one EM-Cell may emulate a cluster of theoretical M-Cells, as long as the projection functions are properly established and the temporal and data (spatial) precisions are satisfactory. This is good news for those who possess large quantities of stored-program computers that are connected to some kind of network or large quantities of processors connected under specific topological structures. However, there are fundamental differences in practical programming methods for stored-program computers and for *xInf Machine*. Programmers now have to add one thing to whatever codes they write: *time*. No matter it is 1 ns, 1 ms, 1 s or 1 hour, every function must contain an apparent value of execution time, not just "i" or "tau", etc. and leave it to the compiler. Precision in physical time is as important or even more important than precision in data (space). Any compiling system and operating system that is not able to 100% guarantee the required timing precision are not allowed to work in the *xInf Machine* system (they can still work for regular *Observers*, though). Moreover, communication devices and protocols that cannot guarantee the timing precision are not permitted, either.

Stored-program computers may be used for constructing M-Cells in a *xInf Machine* if they satisfy the required timing precision. However, they have been designed in very different ways that may not be good enough for this purpose. I have mentioned that FPGA is a good candidate for basic building blocks for M-Cells, so let us see how to go further on this way. It has been quite many years since the initial development of FPGAs and now they are quite extensively used in all kinds of computation and communication devices. However, one common feature of such applications is that each piece of FPGA chip is placed in a circuit board that serves other devices by their protocols, i.e. work as slaves. One of the few occasions when an FPGA chip becomes the master of the computing or communicating device is when it is used as a hardware simulation device for designing ASICs. The ideology behind such a phenomenon is quite simple: in a hardware system where only this piece of component is programmable, who is the one that changes itself in order to serve others? Therefore, my solution is also quite simple: if a hardware system contains nothing but FPGA as computing and communication signal handling components, then every FPGA is a master. This ideology perfectly fits the primary concept of the *xInf Model* that every M-Cell has its own independent *time line*.

So, one just needs to connect some FPGA chips together, but in practice this is not an easy task. The pins on one piece of packaged FPGA chip cannot be easily wired to those on another chip. Because of the high working clock frequency, the wires have to be well placed and proper connector sockets shall be installed, so to avoid electromagnetic interference and mismatch in timing among different signal lines. Furthermore, commercially available FPGA chips usually do not have much on-board memory, if we need functions that do not update at very high repetition rates but must keep large chunks of data over relatively long time intervals, certain size of random-access-memory (RAM) is needed. Moreover, in order to be able to re-write the FPGA diagram (program) while it is working online, there needs to be non-volatile memory that keeps one or multiple backups of the diagrams. Last but not least, the M-Cells still need to communicate with the rest of the digital and physical world via non-FPGA components, so at least one other kind of connection line with proper protocols need to be implemented in M-Cell, for example the PCIe bus that is nowadays standard for desktop computers

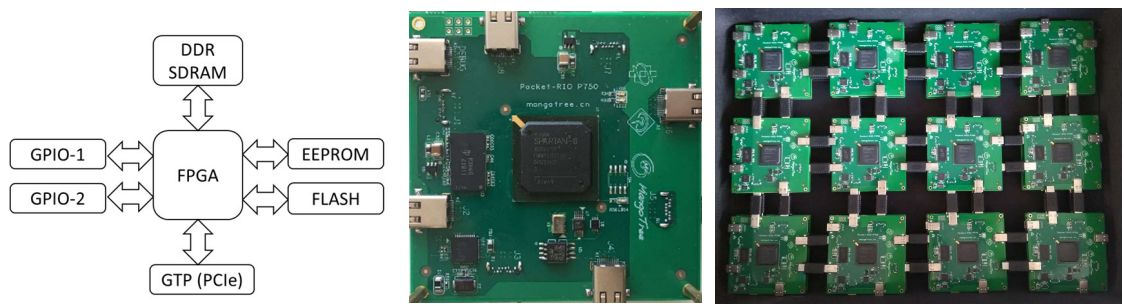

**Fig. 4c.** Example of M-Cell and interconnected matrix. Left: sketch diagram of a M-Cell, the abbreviations on each component means its common meaning in computer engineering, except that the GPIO-1, GPIO-2 stands for multiple general-purpose-input-output (GPIO) lines, effectively these are just simple digital logic lines that satisfies certain standards (e.g. TTL, LVDS, etc.). Middle: in the practical implementation, 128 GPIO lines of a model SPARTAN-6 LX75T FPGA chip were wired to 8 pieces of Type-C USB connector sockets, four on one side and another four sockets on the other side of the base board. The extra Type-C USB is reserved for debug and communication with the compiler PC. This topology of connector allows direct plugging of Type-C USB sockets between two boards and it could also extend in 2D matrix and 3D multi-layers, as shown on the right image.

Taken together, Fig. 4c shows an example of design and realization of such a building block. The unique design of this block is the geometry of the eight 24-pin Type-C connectors (16 digital lines each). Such geometry of the board and connector allows unlimited extension in 2D and 3D space by simple plugging, as shown in Fig. 4c. The way that a 3D connectivity structure is formed by this M-Cell is quite similar to that of the columnar structure of the cerebral cortex. However, if one wants to model the complex cortical connectivity map while each board simulates certain population of neurons, some soft wiring is needed in addition to the hard plugging. This key issue of the xInf architecture is that it allows any arbitrary topology of wiring and input-output configurations, for example a diagram shown in Fig. 4d. Note that the M-Cells do not necessarily connect just to other M-Cells, they can connect to sensors and motor driving modules too, that allows real-time complex feedback control involving multiple sensors (visual, acoustic, mechanic, chemical) and motors, which is highly important for many modern automatic instruments.

Furthermore, there is another possibility that M-Cells work together with conventional computing devices and networks that do not always guarantee real-time operation. The topology has to be restricted to relatively simpler forms, i.e. to hook them onto the same network or host PCIe bus and treat each M-Cell as either a terminal device or hub device in this context, as shown in Fig. 4d. This specific mode of implementation can readily find many applications that involve customized parallel hardware acceleration of certain algorithms, including those commonly used AI algorithms.

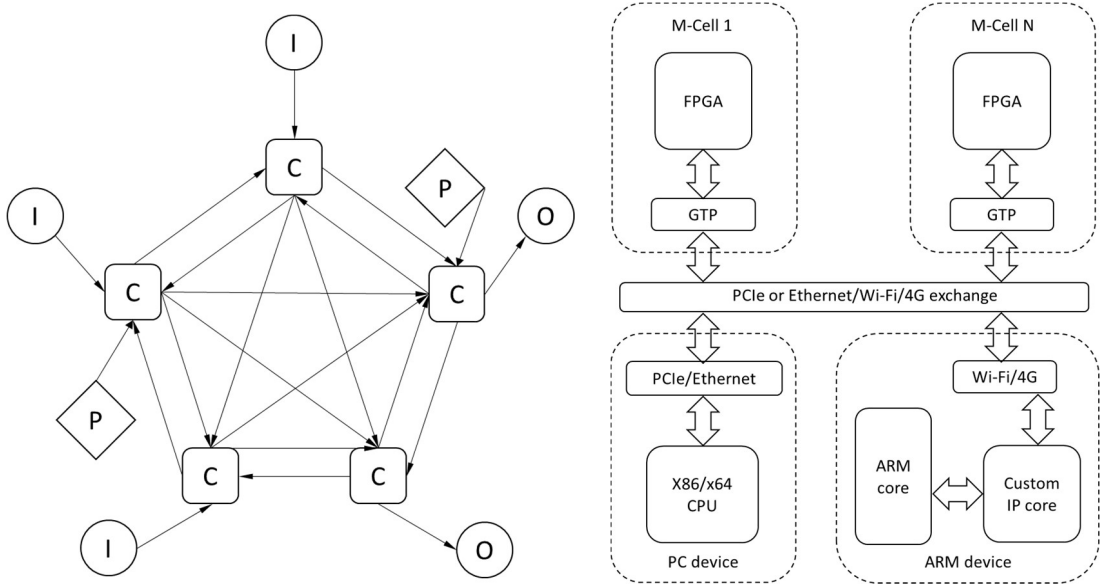

**Fig. 4d.** Topology of the xInf Architecture. Left diagram: when using only *real-time* M-Cells as computational units as well as input/output interfaces, the system connectivity topology can be arbitrary, as an example shown here. Symbols: C for specific computational M-Cells, I for input interface M-Cells, O for output interface M-Cells, and P for power source M-Cells. Right diagram: when using M-Cells together with other computational devices that do not necessarily guarantee real-time operation, the system topology has to comply with those required by the other devices. This example involves a central system bus or network core that distributes the computational tasks and defines request-acknowledgement orders.

Let us conclude this chapter to move on. I have proposed a "All-FPGA" solution for practical implementation of the *xInf Model* and made a specific design and example of basic building blocks. Such a practical architecture has infinite sized *instruction set* if we use the terminology of stored-program computers, in respect to the name of x86 and x64 processors of *Intel*[TM], I gave the name "xInf" for this architecture and its "processor" (there is no actual processor herein). The fact that this architecture enables any arbitrary topology of connectivity while precision in time and data are guaranteed, making it feasible to simulate any physical object in the universe whose *spacetime* topology and functions (transformation, transmission, projection) can be experimentally inferred or determined. Such objects include the natural brain (with mind therein), the research on which largely belongs to the experimental neuroscience.

5. The xInf Model in Neuroscience: quest for the Minimal Complete Set for natural intelligence

This chapter is about just one thing, translation: to re-describe experimental neuroscience by using the language of *xInf Model* in the context of the five sets of rules for engineers and physicists. The translation word chart may increase rapidly when more people are involved to perform meta-analysis on neuroscience and computer engineering literatures. The last four out of the five sets of rules are directly interpretable, whereas the first one, cell birth or death, requires a bit careful definition. The biological notions of birth and death for neurons are slightly different as for other kinds of biological cells, in the sense that growing or losing some extrusion parts (dendrites and axons) of the neurons are also considered to be birth and death, i.e., generation and degeneration. However, in the *xInf language* we shall tell them apart: the creation (or migration) of a new biological cell body (e.g. neuronal somata) or the insertion (or commencing) of a new real-time programmable processor chip is called *birth*, while the dysfunction and/or destruction of biological cell body or the dysfunction (or removal) of a processor chip is called *death*. Partial growth or death of biological cells, as well as re-programming and/or re-wiring on the processor chip, are considered to subject to the second set of rules, i.e. compartmentation and connectivity. In short, the birth and death here mean the appearance or disappearance of an independent *time line*, whatever state space structures comes with this time line are determined in the other sets of rules.

Suppose that experimental neuroscience has already found out as much as possible regarding the five sets of rules, i.e. there is a complete descriptive model of natural brain, why cannot we use stored-program computer to simulate a natural brain? The straight answer is: there is no global causal order or global time. Almost all of the existing artificial neural network models assume that state changes in model neurons happen at the same time (in one iteration), or at least some fraction of the population does so. This is untrue in experimental neuroscience. Two lines of evidences falsify such a presumption. (1) The transduction velocity of action potential along axons[62] as well as dendrites[63] is limited to certain values that are not negligible comparing to the lengths of axons and dendrites in natural neurons in the central nervous system. In contrast, the stored-program computers and network protocols have been deliberately designed to avoid such delay-jitter problems. (2) There is no single or fixed number of clock generators found in natural neural network (whole brain, not just part of it), if there are any, they vary in number, in place and over time, based on neuroscience findings at multiple spatial scales[64]. Suppose all of those experimental findings are true, then a brain must have all these different kinds of different clocks being present, then there is still no global synchronization.

Therefore, every neuron shall be treated individually as a M-Cell. And so? Now that stored-program computers and their networks can also be an EM-Cell that simulates one or multiple M-Cells, and the temporal precision is not that difficult to guarantee if people can use 1 million or even more CPUs in those supercomputers, why cannot we just use stored-program computers to simulate the natural neural system (brain)? The answer is a bit more engineering-wise. In addition to the excitatory neurons and inhibitory neurons[65] that are known to AI modelers[66], there is at least a third category of cells that are not properly represented in known models, i.e. the modulation cells[65,67]. These cells may or may not fire action potentials to directly excite or inhibit other neurons, but they do all kinds of tinkering in the system, e.g., change synaptic efficacy, change transduction velocity, change firing threshold, etc., many of those neuromodulation phenomena and mechanisms have been demonstrated to be causally related to natural behaviors[68-70]. Effectively, many major parameters of neural networks are modulable[67] in real natural neural systems in *real-time*. That is not the worst news except to reduce sleeping time of some programmers and to increase running time of some computers. A worse fact is that: some of these modulatory neurons do receive direct synaptic inputs from normal excitatory and inhibitory neurons by normal neurotransmitters[67], while some receive special molecular messengers from non-neuronal cells and even from outside the brain (please try to recall what happened in the mind when consuming alcoholic drinks, smoking cigarettes, or just being hungry).

This is really troublesome for stored-program computers – where can we store the program of brain beyond the natural body with the brain? My answer is: nowhere.

Fortunately, the fact that natural brain cannot be simulated by a stored-program computer beyond its body does not mean that natural brains (and mind therein) cannot be simulated by non-biological machines. But one thing is for sure: **to simulate a brain (and/or mind therein) there must be one and just that one physical body**. I do not intend to argue to which philosophy this concept belongs, I only propose this concept as a fundamental engineering guideline. This means that if a non-biological machine is about to simulate consciousness mind, it must have sensors and motors that are built together with it in a mechanic body.

Now it is clear why I named the computational unit as M-Cell. Yes, they are machine (non-biological) counterparts of biological cells. More importantly, the way that M-Cells handle digital and analog electronic signals is universal such that sensor devices and motor devices are also considered as M-Cells. With the *xInf Model* we do not only simulate neurons but can also simulate all other kinds of cells as well, i.e. to create a body with brain using non-biological materials. This machine body has its physical shape and is able to autonomously interact with the physical world as much as humans and other biological organisms do. However, what this machine would do in reality is depending on the specific rules of how its M-Cells are working and also depending just on the machine itself, i.e. true individualism.

The comprehensive framework of the *xInf Model* involving these five subsets of rules, are minimally required for engineering because all of them are required for constructing a *Turing Machine* and all computers made by human are some kind of *Turing Machines*. (1) the birth of M-Cells to create the tape cells and the head, note that ideal *Turing Machines* have infinite length of tape so the birth rule guarantees that the machine gets new tape cells whenever needed; (2) the compartmentation and connectivity of M-Cells to establish left-right movement of head over the tape; (3) the transformation functions to define how each tape cell to be written and erased as well as how head state changes; (4) the transmission functions to make the symbols being written to and read from tape cells; (5) the interface functions to read out the symbol(s) from tape when the head state is read out as "halt".

Whilst still feasible by using stored-program computers for implementing the *xInf Model* for reverse engineering of natural intelligence, the FPGA-based approach that I proposed have the following advantages that can well support the five sets of rules:

1) Pure individuality and parallelism, that adding or removing one block does not interfere with the remaining part of the machine except for the necessary connectivity and communication functions to be established. Particularly, adding or removing blocks does not affect the timing mechanism in the remaining blocks.
2) Compartmentations and connections are direct and physical (and visible), no emulation required. If ever emulation comes into concern, it is only for the purpose of constructing an EM-Cell that emulates multiple M-Cells on one piece of chip. The fact that biological neurons do from relatively distinct blocks of tissues (i.e. individual brain regions) makes it easier to translate between neuroscience and engineering in this aspect.
3) Transformation functions execute at specified clock timing with best possible precision, satisfying the three levels of precisions as mentioned in chapter 4.
4) Transmission functions also execute in real-time and the accuracy is guaranteed by oversampling. Transmission requires no encoding and decoding protocols, however, it does allow additional cryptic encoding and decoding in *real-time* to meet communication safety concerns.
5) M-Cells can directly interface with the physical world by means of analog electrical voltage and AD/DA converters without any other digital devices. In fact, many advanced sensor devices and motion control devices are equipped with FPGA with certain firmware for on-board signal processing.

As for how much timing precision is required, I would propose that, in order to faithfully reverse-engineer the natural nervous system instead of running algorithms with arbitrarily defined "time step", there are three levels of the maximum tolerable computational time error. Level 1: 1 ms, which corresponds to typical maximum firing rate (1 kHz) of neurons; Level 2: 0.1 ms, which corresponds to the commonly used sampling rate (10 kHz) for general single-cell electrophysiology experiments; Level 3: 0.01 ms, which corresponds to the very rapid time course of action potentials of some specific types of neurons that are highly sensitive to synaptic input timing due to differences in axon transduction speed.

Since the simulation of mind requires a physical body, the mechanical construction of a machine body is as important as the construction of its electronics circuits. One of the most important method for analyzing and designing engineering materials is the *Finite Element Analysis* (FEA). It has been demonstrated that FPGAs can be used to perform real-time simulations using FEA[71]. This technical approach brings us an important concept, that it is possible that a machine can sense its own body' situation with just a few internal sensors despite that machines cannot easily have thousands of biochemical and biomechanical sensing modalities everywhere in the body as natural biological creatures do.

This manuscript will be uploaded to a pre-publication archive and keep being updated with the latest achievement in the progress of evolution that I show in next chapter. My working hypothesis is that, for each natural biological behavior, there is an underlying **Minimal Complete Set** of rules to be implemented on a *xInf Machine* that can evolve and simulate the behavior. At the moment I cannot tell whether the existing neuroscience knowledge is sufficient for constructing a *Minimal Complete Set* for simulating certain natural intelligence behaviors, neither to prove whether this approach is feasible at all. The only thing that I can do is to translate neuroscience findings to physicists and engineers by re-interpreting publications of all neuroscientists, and try to initiate a new game as discussed in chapter 6.

After all, the ***xInf Model*** **is a universal (*Turing-Complete*) language that can be used to intuitively and strictly translate the knowledge among the fields of Physics, Computer Engineering, and Neuroscience**, as shown in Table 1. The *xInf Model* itself is largely a mathematical language framework involving primarily the **Algebraic Geometry**, it is not the reality, although by wisely designing and building *xInf Machine* in the reality one can simulate many (perhaps all, perhaps not all, I leave this question open) objects in the real universe. With this language I hope to bridge the gaps among physicists, engineers, computer (stored-program) programmers, AI researchers and experimental neuroscientists, and thus to stop arguments like whether mind is quantum computing or what is intelligence. I may as well imitate the famous quote "Shut up and calculate!" from David Mermin that once was misattributed[72] to Richard Feynman, say "Shut up and program!". In this way, perhaps a new job will emerge, i.e., the *xInf Engineers* who has acquired adequate knowledge of physics, material engineering and experimental neuroscience as well as skills in designing and programming distributed real-time computation systems.

| xInf | Math/Physics/Informatics | Computer Engineering | Neuroscience (cellular network level) | Neuroscience (whole-brain level) |
|---|---|---|---|---|
| M-Cell birth & Death | Independent local time dimensions | Add/removing processors with independent clocks | Cell birth & death, identity | Brain regions enable/disable |
| M-Cell compartmentation & connectivity | State space topology | Circuit wiring among processor blocks | Dendrite/axon compartmentation, synaptic connectivity modes for each cell | Circuit map by fMRI, optogenetics, anatomical reconstruction, etc. |
| M-Cell transformation functions | Same definition as M-Cell | Processor instructions on each clock | Membrane potential and synaptic plasticity kinetics for each cell | Neuronal population activity patterns in each brain region |
| M-Cell transmission functions | Same definition as M-Cell | Communication protocols on each processor | Synaptic transmission kinetics for each type of cell | Inter-brain-region projection activity patterns |
| M-Cell – Observer interface functions | Observables and measurement protocols | Input/output protocols on relevant devices | Sensory & motor terminal kinetics for each terminal nerve cells | Sensory organs and muscles' activity patterns |

**Table 1.** Basic translation lookup table for the xInf, math/physics/informatics, computer engineering and neuroscience languages. This is only one startup version of the lookup table, it will keep on evolving by means of more detailed structures and contexts filled in.

**6. From the Imitation Game to the Evolution Game**

Despite the recently appearing "gold rush" in the AI industry, there is very intense debate as for what is the definition of after all, the intelligence. We shall not overlook history that the concept of AI had already been initiated many years ago, when the great mathematician Alan Turing proposed the famous "*Imitation Game*"[73]. The question of whether machines can think could be equivalently transformed as whether machines can imitate human minds.

With the *xInf* (language, model, computational architecture) I propose to start a new game, termed as the *Evolution Game*. The way to play is quite simple as follows. Consider these historically mentioned and upcoming human-proposed tests together as one test – the universal intelligence test, and we know that some biological humans can pass this test. Now, if a non-biological *xInf Machine* cannot pass this test, either or both actions shall be done: (1) By human force, revert the machine to a certain state in history and let it evolve from that time point on once more in different environment; (2) by either human force and/or by the machine itself, change the rules. We can expect (not 100% guaranteed) that within limited number of iterations, the *xInf Machine* may be evolved into a state that it can win the tests. As mentioned in the first chapter, the *xInf Machines* may run in chaotic patterns, however, there are human-specified boundaries and conditions as selection criteria, machines that runs into states that fail to comply with criteria prior to fail in the contest would have been already reset or reconfigured. There will be very likely many humans participating in this game and thus there is a competition, very much like that of the natural history evolution, that is why the game is called the *Evolution Game*.

The rules for constructing such evolutional contestant machines shall be present at multiple spatiotemporal levels. To win in the *Evolution Game*, one shall better not stick to rules just at one level. To illustrate this general guideline, let me use the *StarCraft* game as an example. At the moment it is obvious that the best AI player is still by far worse than human players (https://www.technologyreview.com/s/609242/humans-are-still-better-than-ai-at-starcraftfor-now/). The *StarCraft* game, beyond its real-time challenge, directly involve many human social concepts that are nowadays yet non-modellable with finite order logic. There are two general strategies to get around such a barrier. First way is to combine machine learning algorithms that work very well on localized combat scenes together with conventional weighted decision trees (that are largely determined by humans) for global tactical planning, which is the currently more favorable strategy by big company players. Note: the winners of *StarCraft* AI contests are still individual programmers using simple decision tree script, e.g. using "4/5-Drone Zergling Rush" or other simple tactics, while any trained human *StarCraft* e-Sport player can easily defeat such simple AI tactics. The second way, on which there is not much work yet, is to first build a machine that first learns the cognition of all required concepts and then practices playing the *StarCraft* game. This machine will then be able to play any computer game (may be good at some and not good at others) and shall be regarded as a landmark of truly intelligent machines (and most importantly, having no threat to humans except to those who do e-Sport as professions). The point is, if there are 10000 (literally, many) of such machines, the best one among those is likely to defeat the best human *StarCraft* player under the same physical interface constrains, i.e. the same upper limit of actions-per-minute (APM) and the same visible screen. Note: nowadays all *StarCraft* AI programs cheat in the sense that they see the whole battlefield map with more details than human players.

In this way we convert once more the original question of "can machines think" to "will machines made of M-Cells be eventually able to think". If there could be a positive answer, it would most likely to be in the machines consist of mostly digital M-Cells deal with analog signals of the physical world similar as humans do. We shall also note that, no human is born to be able to think or to calculate, all intelligent abilities are acquired postnatal. Every human and every vertebrate animal, no matter being literally intelligent or not, start growing from (at least) one cell and adapt their cells to the environment in real-time, for all the time and until death.

I must emphasize that this *Evolution Game* has no intention to simulate entire natural brains *per se* no matter being healthy or diseased and it is not a game *in silico*. These machines in this game will not be able to help humans to understand or to cure human neural diseases *per se*, however, given that these machines will be able to autonomously learn and physically interact with natural objects, it is likely some of them could be trained and employed by human scientists to help cure human diseases (and not just neural diseases) by performing caretaking, diagnosis, therapy, surgery, medicine production and inspection, as well as assisting in research experiments in addition to the statistical calculations, image recognitions and numerical simulations.

After all, there is indeed one limiting factor – physical energy. Non-dissipative energy sources are required for both the construction and operation of the M-Cells because the M-Cells are dissipative systems. The amount of materials, on the other hand, is abundant on planet earth, since the major element, silicon, extensively exists in every sand desert. Another good fortune is that the materials required by biological humans and ecology are largely different that those required by semiconductor industry. Moreover, tropical deserts are the most efficient in receiving solar energy. Taken together, the best locations for evolving the society of truly intelligent machines (that are able to build themselves) in the future are the deserts inhabitable for humans.

Nevertheless, this is no longer a scientific or engineering question as for when or how the first truly intelligent machine would be realized. Whether within the limit of available energy on the planet earth the natural intelligent machine could be evolved, as well as whether it should ever be allowed by humans, remains an open debate as well as a challenging question for the entire humankind.

**Acknowledgement**

I would like to hereby thank those who directly helped to the composition of this manuscript: Dr. Arthur Konnerth who over many years had those two-photon microscopes developed and applied for the experimental studies and gave advisory on neuroscience to me, as well as mentioning the keyword of "translation"; Dr. Bert Sakmann who was my other PhD advisor and devised the mouse whisker experiments; Dr. Zsuzsanna Varga who performed the experiment herein described (rehearsed) as well as many other experiments that are essential for the generation of the key concepts in this manuscript, and most importantly, to bring forward and discuss about the nature of space and time; Dr. Xiaowei Chen who had extended the usage of those microscopes and other neuroscience measuring devices built by me and produced important data regarding the perception of time; Dr. Yuguo Tang who is a traditional optical engineer but came to and supported the inspiration and approach on reverse engineering of natural intelligence; Dr. Zhi Jiang who helped on versing the mathematical notations; Dr. Marc-Thierry Jaekel who gave critical view and help on the understanding of physical time; Mr. Xing Wu the engineer who designed the FPGA developer board and brought forward the idea of oversampling for intercommunication.

I would also like to thank my colleagues with whom together we built microscopes and other neuroscience measurement instruments, and particularly thank those neuroscientists who have used those instruments, not for the business part, but for the neuroscience knowledge that I had learnt on those innovative research works, whereas some of them played important inspiring role for this manuscript.

This work has been supported by or involved in the following grants and projects:
At address #1: CAS Pioneer Hundred Talents Program (2016 -), Chinese Academy of Sciences; National Research and Development Projects for Key Scientific Instruments (Project ID: ZDYZ2013-1, 2014-), Chinese Academy of Sciences;
At address #2: "CORTICONIC" FP7-ICT (Project ID: 600806, 2013 - 2016), European Commission; "InVivoSynapse" (Project ID: 323148, 2013 -), European Commission;
At address #3: "973 Program" (Project ID: 2015CB759500, 2015 -), National Natural Science Foundation of China;
At address #4: Beijing Nova Program (2014 - 2017), Beijing Municipal Science and Technology Commission; "HSTR2PM" (Project ID: 31327901, 2014 -), National Natural Science Foundation of China;
At address #5: "U-Labor" Global Lab Sharing Project (2017 -).

**Declaration of Conflicts of Interest**

I hereby certify that this manuscript does not involve any commercial, political, or military interest, that this manuscript is freely accessible on the pre-publication online library "arXiv" https://arxiv.org and "bioRxiv" https://www.biorxiv.org for every individual person and organizations, that the technology described in this manuscript involve patents and patents pending, and that no person or organization is allowed to use or apply the contents of this manuscript for commercial, political or military purposes without written authorization. I also personally retain the rights to develop a non-profiting, non-governmental organization to distribute the serial numbers of M-Cells (engineering definition = individual digital logic devices that have independent clocks, match the diagram of Fig. 4c and use the oversampling method for transmitting data over interconnection digital signal lines) at free charge to every person or organization.